\documentclass{aa}
\usepackage{graphicx}

\usepackage{txfonts} 
\begin{document}

   \title{Serendipitous discovery of the unidentified extended TeV $\gamma$-ray source HESS\,J1303-631}


\author{F. Aharonian\inst{1}
 \and A.G.~Akhperjanian \inst{2}
 \and K.-M.~Aye \inst{3}
 \and A.R.~Bazer-Bachi \inst{4}
 \and M.~Beilicke \inst{5}
 \and W.~Benbow \inst{1}
 \and D.~Berge \inst{1}
 \and P.~Berghaus \inst{6} \thanks{Universit\'e Libre de 
 Bruxelles, Facult\'e des Sciences, Campus de la Plaine, CP230, Boulevard
 du Triomphe, 1050 Bruxelles, Belgium}
 \and K.~Bernl\"ohr \inst{1,7}
 \and C.~Boisson \inst{8}
% ...
}

\institute{
Max-Planck-Institut f\"ur Kernphysik, Heidelberg, Germany
\and
Yerevan Physics Institute, Armenia
\and
University of Durham, Department of Physics, U.K.
\and
Centre d'Etude Spatiale des Rayonnements, CNRS/UPS, Toulouse, France
\and
Universit\"at Hamburg, Institut f\"ur Experimentalphysik, Germany
\and
APC, Paris, France 
\thanks{UMR 7164 (CNRS, Universit\'e Paris VII, CEA, Observatoire de Paris)}
\and
Institut f\"ur Physik, Humboldt-Universit\"at zu Berlin, Germany
\and
LUTH, UMR 8102 du CNRS, Observatoire de Paris, Section de Meudon, France
% ...
}

   \offprints{Matthias Beilicke,\\ \email{matthias.beilicke@desy.de}}

   \date{Received 5 April 2005 / Accepted 09 May 2005}


   \abstract{The serendipitous discovery of an unidentified extended TeV
$\gamma$-ray source close to the galactic plane named
\object{HESS\,J1303-631} at a significance of $21$ standard deviations is reported.
The observations were performed between February and June 2004 with the H.E.S.S.
stereoscopic system of Cherenkov telescopes in Namibia. HESS\,J1303-631 was
discovered roughly $0.6\degr$ north of the binary system
\object{PSR\,B1259-63}/\object{SS\,2883}, the target object of the initial
observation campaign which was also detected at TeV energies in the same field of
view. HESS\,J1303-631 is extended with a width of an assumed intrinsic Gaussian
emission profile of $\sigma = (0.16 \pm 0.02) \, \degr$ and the integral flux above
$380\,\mathrm{GeV}$ is compatible with constant emission over the entire
observational period of $(17 \pm 3)\, \%$ of the Crab Nebula flux. The measured
energy spectrum can be described by a power-law $\mathrm{d}N/\mathrm{d}E \sim
E^{-\Gamma}$ with a photon index of $\Gamma = 2.44 \pm 0.05_{\mathrm{stat}} \pm
0.2_{\mathrm{syst}}$. Up to now, no counterpart at other wavelengths is identified.
Various possible TeV production scenarios are discussed.


   }

   \titlerunning{Detection of the Unidentified TeV Source HESS\,J1303-631}
   \authorrunning{Aharonian et al.}

   \maketitle

%
\section{Introduction}

Cherenkov telescopes of ground based $\gamma$-ray astronomy achieve
a very high sensitivity. Due to the limited field of view the
pointing of the telescopes has in general to be decided on the basis of
observations in other wavelengths. However, the comparatively large
field of view of instruments such as HEGRA ($4.3\degr$ full angle) and
H.E.S.S. ($5\degr$ full angle) together with the stereoscopic
observation mode allows nevertheless a scan of a part of the sky
(Aharonian et al. \cite{HEGRA_GalPlaneScan, HESS_GalPlaneScan}) and in
particular allows a search for unknown sources in the field of view of
individual pointings.  In this way the first unidentified TeV
$\gamma$-ray source TEV J2032+4130 has been discovered by HEGRA
(Aharonian et al. \cite{Aharonian:Cyg1, Aharonian:Cyg2}) in archival
data and now, with the H.E.S.S. telescopes, the second unidentified
source HESS\,J1303-631 was discovered, followed by more unidentified sources found in 
the H.E.S.S. galactic plane scan (Aharonian et al. \cite{HESS_GalPlaneScan}). The H.E.S.S. 
telescopes have originally been directed to search for TeV $\gamma$-ray emission from
the binary system PSR\,B1259-63/SS\,2883 near periastron beginning in
February 2004. This binary system has been detected at TeV energies
(Beilicke et al. \cite{HESS_PSRB1259_Telegramm}) which is the subject
of a parallel paper (see Aharonian et al. 
\cite{HESS_PSRB1259_Detection}).  Surprisingly, another TeV
$\gamma$-ray source located at a position roughly $0.6\degr$ north of
the position of the binary system was discovered in the same field of
view, given the name of HESS\,J1303-631 (see Fig.~\ref{FIG:SkyMap}).
The detection and basic features of this new source as well as a search
for possible counterparts in other wavelengths and discussion of
possible TeV $\gamma$-ray production scenarios are reported in this
paper.

%
\section{The H.E.S.S. Cherenkov telescopes}

The H.E.S.S. (High Energy Stereoscopic System) collaboration operates an
array of four imaging atmospheric Cherenkov telescopes (IACTs) optimised
for an energy range of $\gamma$-rays between $100 \, \mathrm{GeV}$ and
$20\, \mathrm{TeV}$. The telescopes are located in the Khomas Highland in
Namibia ($23\degr 16\arcmin 18\arcsec \, \mbox{S}$, $16\degr 30\arcmin
1\arcsec \, \mbox{E}$) at a height of $1\,800 \, \mbox{m}$ above sea
level. Each telescope has a $107\,\mbox{m}^{2}$ tessellated mirror surface
(Bernl\"ohr et al. \cite{HESS_Mirror1}; Cornils et al.
\cite{HESS_Mirror2}) and is equipped with a 960 photomultiplier tube
camera with a field of view diameter of $\sim 5\degr$ (Vincent et al.
\cite{HESS_Camera}) which allows searches and studies of TeV $\gamma$-ray
sources in sky regions of more than $3\degr \times 3\degr$ per pointing.
The telescopes are operated in a coincident mode (Funk et al.
\cite{HESS_Trigger}) assuring that an event is always recorded by at least
two of the four telescopes allowing for stereoscopic reconstruction of the
shower parameters. More information about H.E.S.S. can be found in Hinton
(\cite{HESS_Status}).

%
\section{Dataset}

\begin{figure*}[!t]
   \centering
	\resizebox{\hsize}{!}{
	\includegraphics{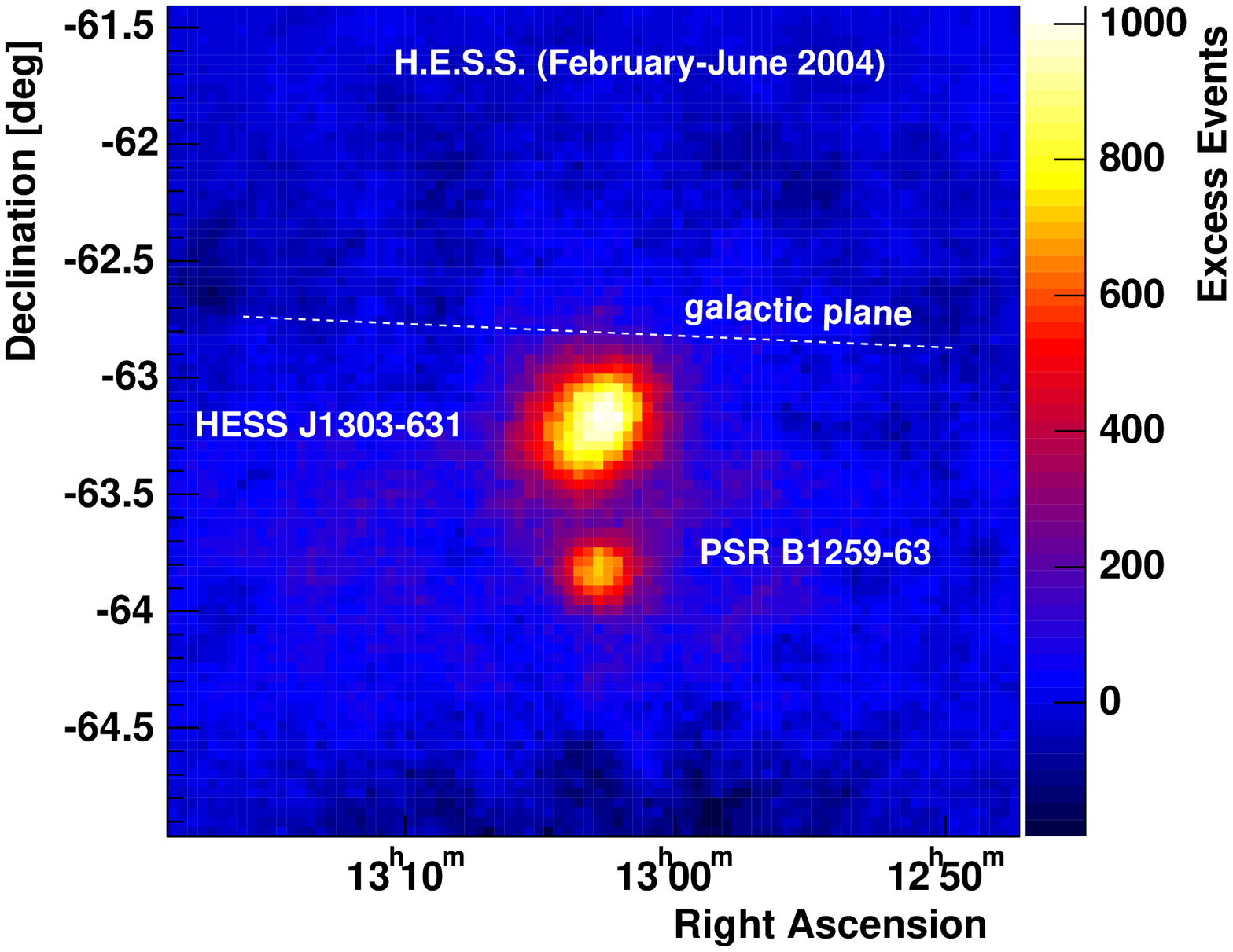}
	\includegraphics{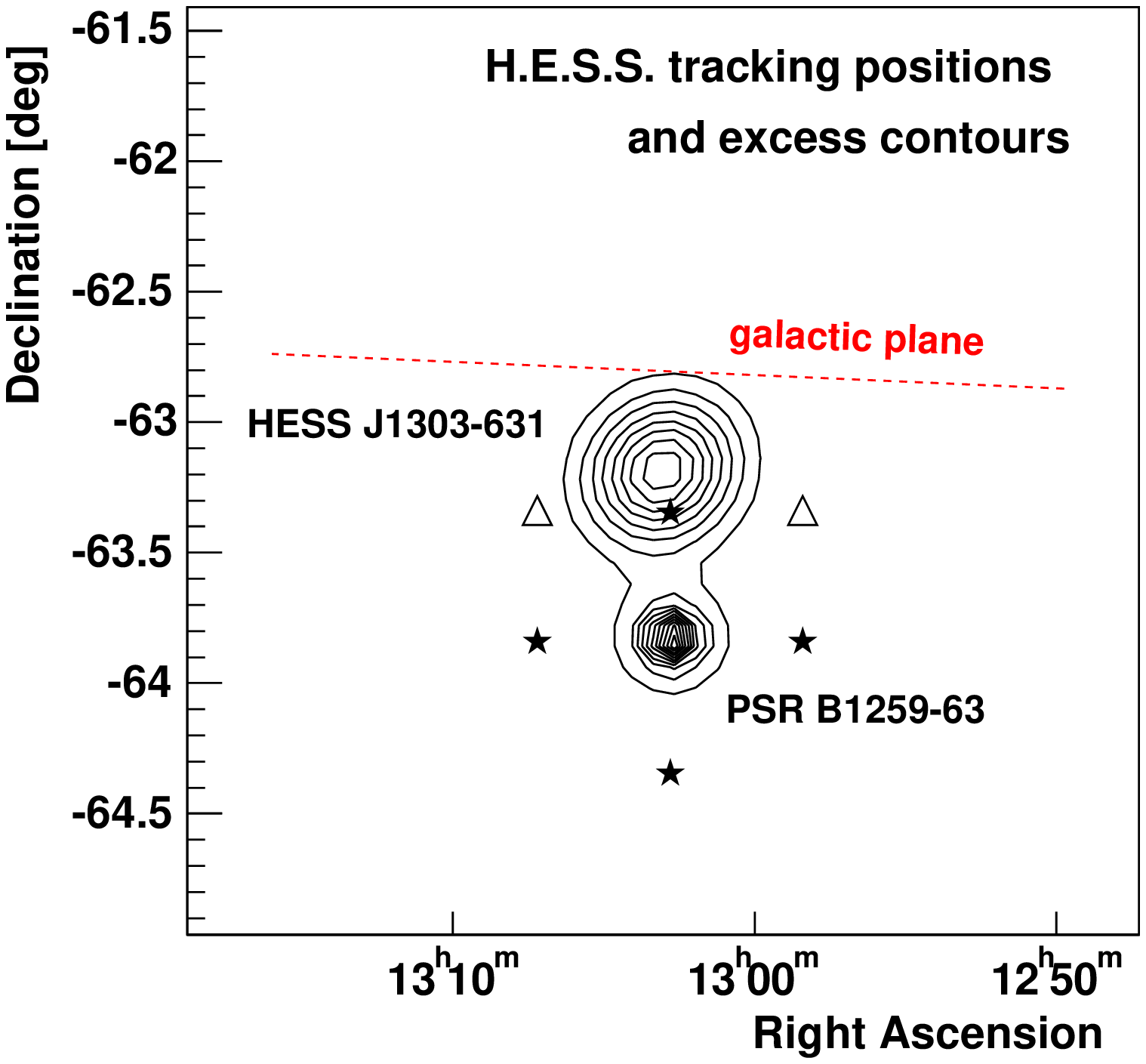}
         }
  \caption{\emph{Left:} The sky map showing both TeV $\gamma$-ray sources:  
HESS\,J1303-631 and PSR\,B1259-63/SS\,2883. The Galactic plane is also
indicated. The number of events are integrated within a circle of $\Theta 
\le 0.14\degr$ for each of the correlated bins. \emph{Right:} The 
function which was fitted to the (uncorrelated) excess sky map is drawn 
as contour lines. Also shown are the different H.E.S.S. tracking 
positions: Filled stars indicate the initial tracking positions (February 
- May) and the empty triangles the optimised positions which were used 
since May, 2004.}
  \label{FIG:SkyMap}
\end{figure*}

The data were taken between February and June 2004 with the fully
operational H.E.S.S. IACT array. The average zenith angle of the
observations was $42.7\degr$ yielding an energy threshold of
$E_{\mathrm{thr}} = 380 \,\mbox{GeV}$, defined by the peak
$\gamma$-ray detection rate of a Crab-like spectrum after event
selection cuts. The observations were performed in the \emph{wobble}
mode, tracking a position shifted by $\pm 0.5\degr$ in Declination or
Right Ascension with respect to the nominal source position (in this
case the PSR\,B1259-63/SS\,2883 position of the initial observation
campaign), allowing for an unbiased simultaneous background
determination for not too extended sources.  Following the detection
of HESS\,J1303-631 in the PSR\,B1259-63/SS\,2883 field of view, the
telescope tracking positions were changed to new sky positions in May
2004, optimised for both sources by choosing the wobble modes with
respect to a position located between PSR\,B1259-63/SS\,2883 and
HESS\,J1303-631. In the right hand panel of Fig.~\ref{FIG:SkyMap} the
initial pointing positions (filled stars) as well as the new pointing
positions (empty triangles) are shown.

The data were selected by standard quality criteria (stable weather and detector
status) leaving $54.5$ hours of data ($48.6\,\mbox{h}$ detector life time) for the
final analysis. For the data taken between February 26 and March 5, 2004
($7.6$ hours), one of the four telescopes was excluded from the analysis due
to technical reasons with the event read-out of the telescope. The remaining
data (March to June 2004) were analyzed using the full array of four telescopes. The
raw data were subject to the standard calibration (Aharonian et al.
\cite{HESS_Calibration}) and Hillas parameter-based analysis (see Aharonian et al.
\cite{HESS_Analyse} for more details).

\section{Analysis and Results}

\subsection{Discovery of TeV $\gamma$-ray emission from HESS\,J1303-631}

Shortly after the discovery of TeV $\gamma$-ray emission from the binary
system PSR\,B1259-63/SS\,2883 the highly significant excess of the second
source HESS\,J1303-631 was found in the data. The discovery sky map
resolving both TeV sources is shown in the left hand panel of
Fig.~\ref{FIG:SkyMap}. The number of excess events was calculated using
the ring background model in which the background is determined from a
ring region with a radius $r > 0.5\degr$ centred around the putative
source position. Both TeV $\gamma$-ray source regions were excluded from
the background region to avoid background contamination by real
$\gamma$-ray events. In order to improve the angular resolution for the
investigation of the position and extension, for the sky map at least
three images instead of two were required per event.

\subsection{Consistency checks}

Since with the discovery of HESS\,J1303-631 for the first time in the H.E.S.S.  
data and in TeV $\gamma$-ray astronomy in general, two sources were found in the
same field of view\footnote{Meanwhile, two sources in one field of view have also
been observed in other pointings with H.E.S.S., e.g. in the galactic center region
and the galactic plane scan.} a wide range of consistency checks was applied and
successfully passed underlining the celestial origin of the measured excess. For
this purpose the data were divided into different subsets and the results (source
strength, excess position, etc.) were compared. The subsets were chosen according to
the different observation periods (from February to June, 2004) and the different
telescope tracking positions (compare Fig.\ref{FIG:SkyMap}, right). Different cuts
on the telescope multiplicity $m$ ($m \ge 2$, $m = 2$, $m \ge 3$, $m = 3$ and $m =
4$) for individual events as well as different cuts on the image amplitude were
applied to the data. Also, individual telescopes in turn were excluded from the
analysis. In all these checks the excess behaved as a genuine $\gamma$-ray source;
more details can be found in Beilicke et al. (\cite{HDGS2004}).

\begin{figure}
  \resizebox{\hsize}{!}{\includegraphics{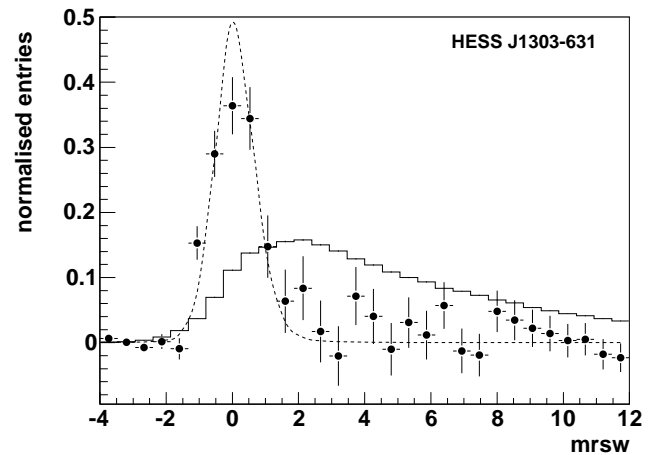}}
  \caption{Distribution of the \emph{mean reduced scaled width (MRSW)} parameter (see text) for 
the HESS\,J1303-631
excess (data points). For comparison, the distributions for simulated 
$\gamma$-ray events
(dashed line) and for hadronic events (histogram) are also shown. The excess measured
from HESS\,J1303-631 is compatible with $\gamma$-ray events.}
  \label{FIG:mscw}
\end{figure}

Finally, an important check concerns the distribution of the
\emph{mean reduced scaled width (MRSW)} parameter which is used for
the $\gamma$-hadron separation (see Aharonian et al.
\cite{HESS_Analyse}). The \emph{MRSW} is obtained by averaging the
reduced scaled widths of the individual images of an event. The
reduced scaled width is defined as the difference (measured in
standard deviations) between the measured and the expected width of an
image. The expected width is obtained from Monte Carlo simulations for
$\gamma$-ray showers, being parameterized as a function of the image
amplitude, the shower impact parameter, and the zenith angle. The
distribution of the \emph{MRSW} parameter of the HESS\,J1303-631
excess is shown in Fig.~\ref{FIG:mscw} together with the distributions
for $\gamma$-ray events from simulations and for hadronic events from
a background measurement. The \emph{MRSW} distribution of
HESS\,J1303-631 is compatible with $\gamma$-ray events.

All results obtained from the different investigations are in good
agreement. They were cross-checked with the template background model
(Rowell \cite{TemplateModel}) and were also confirmed by an alternative
model analysis technique (de~Naurois et al. \cite{HESS_MODEL}). The
celestial origin of the measured excess from HESS\,J1303-631 can therefore
be considered as established.

\begin{figure}
  \resizebox{\hsize}{!}{\includegraphics{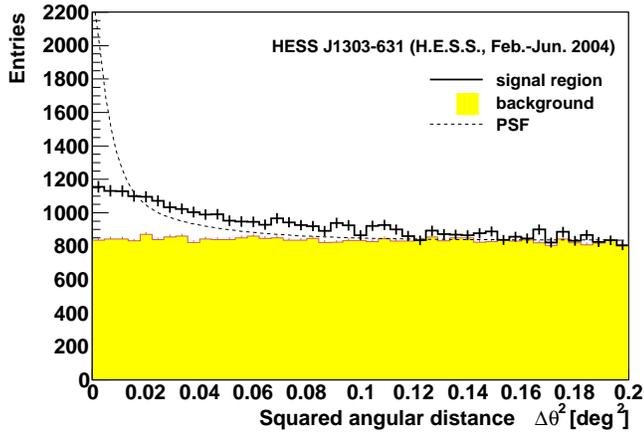}}
  \caption{Distribution of ON-source events (solid histogram) and
normalized OFF-source events (filled histogram) vs. the squared angular
distance $\Delta\Theta^{2}$ between the reconstructed shower direction and
the object position. The expected ON-distribution for a point-like
source (TeV point spread function, PSF) is indicated by the dashed line; it is
normalized to give the same number of excess events as HESS\,J1303-631.}
  \label{FIG:OnOff}
\end{figure}

\subsection{Excess and sky position}

To obtain the position of HESS\,J1303-631 a function describing the
excess of the two sources in the field of view was fitted to the
uncorrelated sky map which was generated with a cut on at least three
images per event in order to obtain a better angular resolution. The
excess of HESS\,J1303-631 was fitted by a 2D elliptical function
(two-dimensional Gaussian function with a $\sigma$ for the Right
Ascension and one for the Declination as well as a covariance).
Simultaneously, the PSR\,B1259-63/SS\,2883 position was fitted by the
H.E.S.S. point spread function (PSF) for TeV $\gamma$-rays to avoid a
systematic influence on the fit by excess events from PSR\,B1259-63
leaking into the HESS\,J1303-631 region. The PSF can be described by a
2D double Gaussian function $a \cdot \exp ( \frac{\Delta \bf{r}^{2}}{2
\sigma^{2}_{1}} ) + b \cdot \exp ( \frac{\Delta \bf{r}^{2}}{2
\sigma^{2}_{2}} )$ which was derived from Monte Carlo simulations. The
fit range covers a region of $2\degr \times 2\degr$ centred at the
HESS\,J1303-631 position. The parameters of the PSF ($\sigma_{1}$,
$\sigma_{2}$ and $a/b$) which was applied for the position of
PSR\,B1259/SS\,2883 were left free. The contour lines of the function
are shown in the right hand panel of Fig.~\ref{FIG:SkyMap}. The
position of HESS\,J1303-631 was found to be $\alpha = 13^{\mathrm{h}}
03^{\mathrm{m}} 0\fs4 \pm 4\fs4$ and $\delta = -63\degr 11\arcmin
55\arcsec \pm 31\arcsec$ (J2000.0). The length and width of the fitted
ellipse are within errors the same and clearly larger than that
expected for a point source (see below). Therefore, the
HESS\,J1303-631 excess is compatible with an extended and rotationally
symmetric structure. The position of PSR\,B1259-63/SS\,2883 obtained
from the same fit was found to be $\alpha_{\mathrm{1259}} =
13^{\mathrm{h}} 02^{\mathrm{m}} 49\fs2 \pm 3\fs6$ and
$\delta_{\mathrm{1259}} = -63\degr 50\arcmin 2\farcs4 \pm 21\farcs1$
which is in good agreement with its nominal position of
$\alpha_{\mathrm{1259, nom}} = 13^{\mathrm{h}} 02^{\mathrm{m}} 47\fs7$
and $\delta_{\mathrm{1259, nom}} = -63\degr 50\arcmin 8\farcs8$. The
width of the PSR\,B1259-63/SS\,2883 excess was found to be compatible
with the width of a point-source (Aharonian et al.
\cite{HESS_PSRB1259_Detection}). Fixing the position of
PSR\,B1259-63/SS\,2883 in the fit to its nominal sky coordinates does
not change the results for the HESS\,J1303-631 position and extension
within statistics. The systematic pointing uncertainty of the H.E.S.S.
telescopes is estimated to be $\sim 20\arcsec$ for Right Ascension and
Declination.

The distribution of the number of events in squared angular distance
$\Delta\Theta^{2}$ measured between the reconstructed shower direction and
the derived HESS\,J1303-631 position is shown in Fig.~\ref{FIG:OnOff}. The
PSF of the H.E.S.S. detector for TeV $\gamma$-rays is also shown (dotted
line). A single Gaussian function describing the intrinsic source profile
of HESS\,J1303-631 was folded with the PSF. The folded function was fitted
to the excess distribution resulting in an intrinsic width of
$\sigma_{\mathrm{HESSJ1303}} = ( 0.16 \pm 0.02) \, \degr$ with a
$\chi^{2}$/d.o.f. $= 21/42$.

Adjusting the angular cut to the derived extension ($\Delta\Theta^{2}
\approx \Delta\Theta^{2}_{\mathrm{pointsrc}} +
\sigma^{2}_{\mathrm{HESSJ1303}} \approx 0.05 \, \mathrm{deg}^{2}$) one 
obtains for the HESS\,J1303-631 position $2469 \pm 119$ excess events
corresponding to a significance of $21$ standard deviations calculated
following Li \& Ma (\cite{LiMa}).

\begin{figure}
  \resizebox{\hsize}{!}{\includegraphics{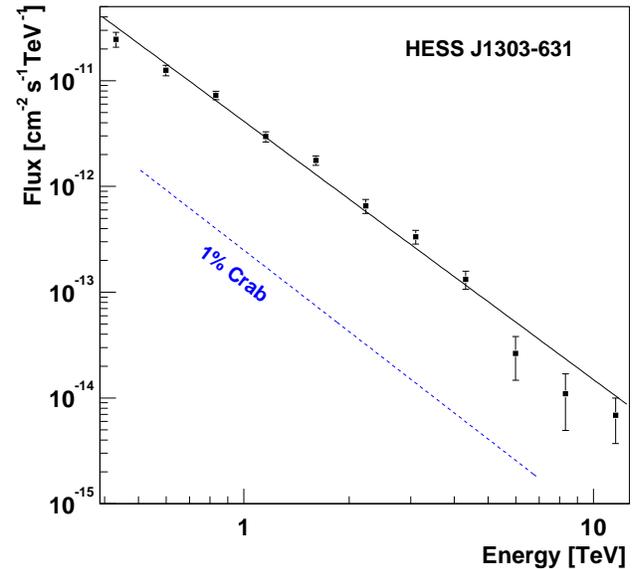}}
  \caption{The differential energy spectrum of HESS\,J1303-631. A fit to 
the data with a power law (solid line) is 
shown. The power-law corresponding to a spectrum of $1\,\%$ of the 
Crab Nebula is also indicated (dashed line).}
  \label{FIG:Spectrum}
\end{figure}

\begin{figure}
  \resizebox{\hsize}{!}{\includegraphics{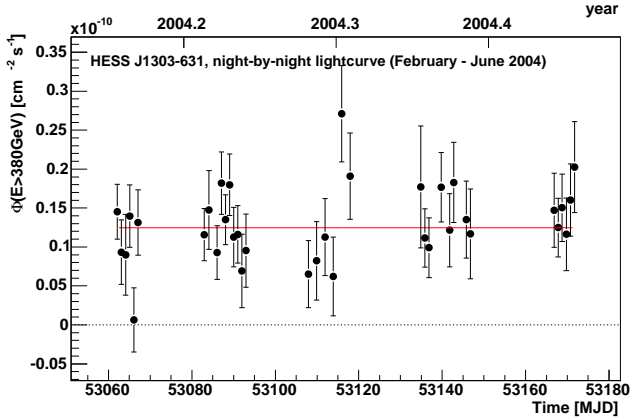}}
  \caption{The HESS\,J1303-631 light curve covering February until June,
2004. Shown is the integral flux $\Phi(E>380 \, \mathrm{GeV})$ vs. time in
night-by-night bins. The fit of a constant function (solid line) results 
in a $\chi^{2}$/d.o.f. of $35/35$ being compatible with constant 
emission.}
  \label{FIG:LightCurve}
\end{figure}

\subsection{Energy spectrum and light curve}

An energy spectrum was derived using the angular cut of
$\Delta\Theta^{2} = 0.05 \, \mathrm{deg}^{2}$. To obtain the full flux
integrated over the whole emission region of HESS\,J1303-631 the
flux normalisation was corrected for the derived source extension
(assuming an intrinsic Gaussian emission profile). Since this
correction depends on the exact -- and possibly energy dependent --
shape of the source, it introduces a systematic error. Together with
the effects of atmospheric extinction variations and energy
calibration of the detector the systematic error on the flux was
estimated to be in the order of $\sim 30\,\%$.

The spectrum is shown in Fig.~\ref{FIG:Spectrum}. It was fitted by a power-law
$\mathrm{d}N/\mathrm{d}E = N_{0} \cdot (E/1\,\mbox{TeV})^{-\Gamma}$ with a resulting
photon index of $\Gamma = 2.44 \pm 0.05_{\mathrm{stat}} \pm 0.2_{\mathrm{syst}}$ and
a normalization of $N_{0} = (4.3 \pm 0.3_{\mathrm{stat}}) \cdot 10^{-12} \,
\mathrm{cm}^{-2} \, \mathrm{s}^{-1} \, \mathrm{TeV}^{-1}$ with a $\chi^{2}$/d.o.f.
of $27/9$.The integral flux above $380 \, \mathrm{GeV}$ was calculated to be $\Phi(E
> 380\,\mathrm{GeV}) = (1.2 \pm 0.2_{\mathrm{stat}})  \cdot 10^{-11} \,
\mathrm{cm}^{-2} \, \mathrm{s}^{-1}$ corresponding to $(17 \pm 3) \, \%$ of the flux
of the Crab Nebula. Spectra obtained from the different tracking positions as well
as from the different observation periods were found to be compatible within
statistical errors. A fit of a curved spectral shape, e.g. with a power-law
and a cut-off, yields a better $\chi^{2}$. However, at the present understanding of
the systematic effects for extended sources observed with asymmetric wobble
positions (comp. Fig.~\ref{FIG:SkyMap}, right), we prefer not to quote numbers at
this stage but to leave it to more detailed studies on morphology and
spatially-resolved energy spectra which are underway.

A light curve of the integral flux $\Phi(E > 380\,\mathrm{GeV})$ was
derived on a night-by-night basis and is shown in
Fig.~\ref{FIG:LightCurve}. To obtain the integral flux, the count rates
for each night were corrected using the effective areas and an assumed
power-law as obtained from the overall differential energy spectrum. A fit
of a constant function to the light curve results in a $\chi^{2}$/d.o.f.  
of $35/35$ and therefore indicates constant emission from HESS\,J1303-631
during February until June, 2004.

%
\section{Search for possible counterparts}

\subsection{Search in catalogues}

\begin{figure}
\resizebox{\hsize}{!}{\includegraphics{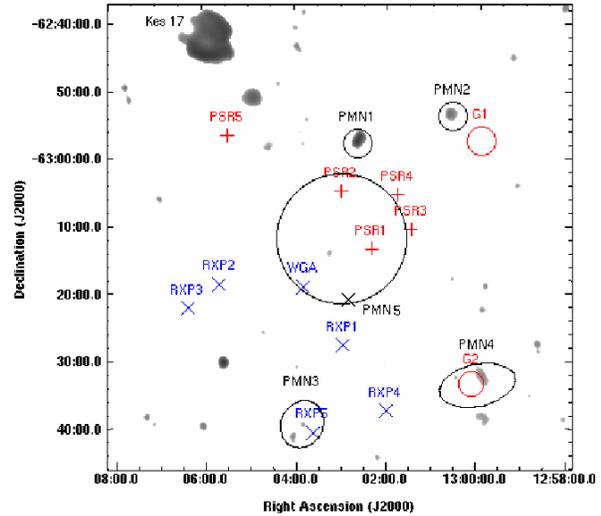}}
\caption{843\,MHz SUMSS radio map of the region around
HESS\,J1303-631 (marked as the large circle of radius $0.16\degr$).
Also marked are the positions of the X-ray sources (WGA and RXP1-5) as well as
the radio sources (PMN1-5, G1 and G2 for \ion{H}{ii} regions and PSR1-5 for pulsars)
listed in Table~\ref{CounterParts}. Circles or ellipses surrounding individual
sources indicate the catalogued source extension (convolved with the relevant
radio beam size).}
\label{FIG:Counterparts}
\end{figure}

\begin{table}[h]
\caption[]{X-ray (ROSAT) and radio sources at a distance of less than $30'$
from the centre of HESS\,J1303-631. The radio sources marked PMN are from the
Parkes-MIT-NRAO catalogue, those marked G are \ion{H}{ii} regions, and those
marked PSR are from the ATNF pulsar catalogue. Shown
are the source name, the ID used in Fig.~\ref{FIG:Counterparts}, the distance
to the HESS\,J1303-631 centroid and the J2000.0 source coordinates.}
\label{CounterParts}
\centering {\scriptsize
\begin{tabular}{llrrr}
\hline \hline
\noalign{\smallskip}
source name & ID & dist. & RA & Dec \\
\noalign{\smallskip}
\hline
\noalign{\smallskip}
\multicolumn{3}{l}{\quad \quad \quad X-ray sources (ROSAT)} \\
1WGA\,J1303.8-6319     & WGA  &  $9.1'$ & $13\,03\,50.9$ & -$63\,19\,06$ \\
2RXP\,J130257.9-632750 & RXP1 & $15.9'$ & $13\,02\,57.9$ & -$63\,27\,50$ \\
2RXP\,J130540.5-631839 & RXP2 & $19.2'$ & $13\,05\,40.5$ & -$63\,18\,39$ \\
2RXP\,J130622.7-632210 & RXP3 & $24.9'$ & $13\,06\,22.7$ & -$63\,22\,10$ \\
2RXP\,J130200.4-633730 & RXP4 & $26.4'$ & $13\,02\,00.4$ & -$63\,37\,30$ \\
2RXP\,J130337.8-634045 & RXP5 & $29.1'$ & $13\,03\,37.8$ & -$63\,40\,45$ \\
\noalign{\smallskip}
\multicolumn{3}{l}{\quad \quad \quad radio sources} \\
PMN\,J1302-6257 & PMN1 & $14.3'$ & $13\,02\,38.9$  & -$62\,57\,48$ \\
PMN\,J1300-6253 & PMN2 & $24.6'$ & $13\,00\,34.4$  & -$62\,53\,43$ \\
PMN\,J1303-6339 & PMN3 & $28.2'$ & $13\,03\,52.6$  & -$63\,39\,29$ \\
PMN\,J1259-6333 & PMN4 & $29.7'$ & $12\,59\,59.1$  & -$63\,33\,37$ \\
PMN\,J1302-6321 & PMN5 &  $9.4'$ & $13\,02\,56.7$  & -$63\,21\,20$ \\
\noalign{\smallskip}
\multicolumn{3}{l}{\quad \quad \quad \ion{H}{ii} regions} \\
G\,303.9-0.1    & G1   & $25.4'$ & $12\,59\,56.6$  & -$62\,57\,22$   \\
G\,303.9-0.7    & G2   & $28.9'$ & $13\,00\,07.3$  & -$63\,33\,20$   \\
\noalign{\smallskip}
\multicolumn{3}{l}{\quad \quad \quad pulsars} \\
PSR\,J1302-6313 & PSR1 &  $4.9'$ & $13\,02\,19.2$  & -$63\,13\,29$   \\
PSR\,J1303-6305 & PSR2 &  $6.9'$ & $13\,03\,00.0$ & -$63\,05\,01$ \\
PSR\,J1301-6310 & PSR3 & $10.4'$ & $13\,01\,28.3$ & -$63\,10\,41$ \\
PSR\,J1301-6305 & PSR4 & $10.6'$ & $13\,01\,45.8$ & -$63\,05\,34$ \\
PSR\,J1305-6256 & PSR5 & $22.6'$ & $13\,05\,28.0$  & -$63\,56\,39$   \\
\noalign{\smallskip}
\hline
\end{tabular}
}
\end{table}

As HESS\,J1303-631 is a new TeV source, a search for possible counterparts
at other wavelengths was performed, using the VizieR service (Ochsenbein
et al. \cite{Ochsenbein}) at the Centre de Donn\'ees astronomiques de
Strasbourg. The closest EGRET ($>100\,\mbox{MeV}$) source,
3EG\,J1308--6112, has its centroid at an angular distance of $2.1\degr$
($\mathrm{RA} = 197\fdg18$, $\mathrm{Dec} = -61\fdg22$), with a nominal
$95\,\%$ confidence radius of $0.71\degr$ (Hartman et al.\
\cite{Hartman}). This source is flagged as possibly extended or multiple
and confused, but examination of the actual position confidence contours
confirms that it can be ruled out as a counterpart.

In the X-ray domain, the only catalogued sources found with VizieR within
$30'$ of the centroid of HESS\,J1303-631 are from the Second ROSAT ($0.24
- 2.0\,\mbox{keV}$) Source Catalogue of Pointed PSPC Observations (5
sources) or its WGACAT version (1 additional source). These ROSAT
sources are listed in Table~\ref{CounterParts}, and their positions are 
marked in Fig.~\ref{FIG:Counterparts}. The concentration of these sources 
to the South of HESS\,J1303-631 reflects the non-uniform sensitivity of 
PSPC observations across this field. Dedicated X-ray observations will be
necessary to make a more complete search for counterparts.

As HESS\,J1303-631 is within the Galactic plane, optical or infrared
sources are too numerous and have not been examined.

To search for counterparts in the radio domain, the 843\,MHz map of the
region around HESS\,J1303-631 was obtained from SUMSS survey (Bock et al.\
\cite{Bock}) and is shown as a grey-scale image in
Fig.~\ref{FIG:Counterparts}. Those relatively bright sources within $30'$
of HESS\,J1303-631 which are also identified in the $4.85\,\mbox{GHz}$ PMN
(Parkes-MIT-NRAO) survey catalogue (Wright et al.\ \cite{Wright}) are
marked as circles or ellipses\footnote{The objects marked PMN1 and PMN2
are consistent with point sources, and their diameter reflects the PMN
beam size, while PMN3 and PMN4 are flagged as extended.} in
Fig.~\ref{FIG:Counterparts} and are listed in Table~\ref{CounterParts}
along with PMN\,J1302-6321 which is the only catalogue source within
the 1-$\sigma$ source extension of HESS\,J1303-631, but is not detected in
the 843\,MHz map. The region overlapping the centre of HESS\,J1303-631
appears devoid of conspicuous radio sources.

The nearest catalogued supernova remnant, Kes\,17 (G\,304.6+0.1), lies at
a distance of $36'$ from HESS\,J1303-631 (Green \cite{Green}). VizieR also
finds two \ion{H}{ii} regions within a $30'$ radius, G\,303.9-0.1 and
G\,303.9-0.7 (Paladini et al.\ \cite{Paladini}). Finally, version 1.21 of
the ATNF (Australia Telescope National Facility) Pulsar
Catalogue\footnote{{\tt http://www.atnf.csiro.au/research/pulsar/psrcat}}
(Manchester et al.\ \cite{Manchester}) lists five pulsars within a $30'$
radius around HESS\,J1303-631. The most compelling positional coincidences
of radio sources with HESS\,J1303-631 appear to be with some of the
pulsars, discussed below.

\subsection{Possible pulsar wind nebula associations}

Pulsars can have associated very high energy (VHE) $\gamma$-ray emission
from the nebulae of high-energy particles (mostly electrons and
positrons) which they produce, as the well-known example of the Crab
Nebula shows. The TeV radiation is thought to arise from inverse Compton
(IC) scattering of lower-energy ambient photons by these energetic
electrons, which also emit synchrotron radiation at lower energies,
typically between the infrared and X-ray bands.  In the case of a
relatively young (950 year old) pulsar like the Crab, the pulsar wind
nebula (PWN) is fairly compact and appears point-like in VHE
$\gamma$-rays, but in older pulsars the emission might be considerably
more extended, due to the expansion of the PWN itself (e.g. van der Swaluw
et al. \cite{vanderSwaluw}) or due to the escape of the high-energy
particles out of the synchrotron-emitting PWN, as proposed by Aharonian et
al. (\cite{AhaAtoKif}).

\begin{table}
\caption[]{Parameters of the pulsars possibly associated
with HESS\,J1303-631. Listed are the pulsar name, its period $P$,
estimated distance $D$, spindown age $\tau$ and flux
$F_{\mathrm sd}$ (see text).}
\label{PulsarParameters}
\centering
\begin{tabular}{llrrr}
\hline \hline
\noalign{\smallskip}
PSR name & $P$ (s) & $D$ (kpc) & $\tau$ (kyr) &
   $F_{\mathrm sd}$ (erg\,cm$^{-2}$\,s$^{-1}$) \\
\noalign{\smallskip}
\hline
\noalign{\smallskip}
J1302-6313 & 0.968 & 28.06 &    2420 & $2.9 \times 10^{-15}$ \\
J1303-6305 & 2.307 & 13.62 & 16\,800 & $3.1 \times 10^{-16}$ \\
J1301-6310 & 0.664 &  2.06 &     186 & $1.5 \times 10^{-11}$ \\
J1301-6305 & 0.185 & 15.84 &      11 & $5.6 \times 10^{-11}$ \\
J1305-6256 & 0.478 & 30.00 &    3590 & $7.1 \times 10^{-15}$ \\
\noalign{\smallskip}
\hline
\end{tabular}
\end{table}

To examine the possibility that HESS\,J1303-631 might be associated
with such an extended nebula from one of the positionally compatible
pulsars, the latters' energetics are now considered. The total
available energy output of a pulsar is given by its spindown
luminosity $\dot{E}$. It is believed that the major fraction of this
energy is ultimately converted to accelerated particles, and the
energetic electrons and positrons can radiate their energy away
efficiently in the form of synchrotron and inverse Compton emission.  
In such a scenario the measured VHE energy flux from a source at the
distance $D$ can be compared to $F_{\mathrm{sd}} \equiv \dot{E}/(4\pi
D^2)$, whereby the conversion efficiency is expected to be much less 
than 100\%, i.e. the measured VHE flux is expected to be much less 
than $F_{\mathrm{sd}}$.

Table~\ref{PulsarParameters} lists the relevant parameters for the
five pulsars included in Table~\ref{CounterParts}, as derived from
version 1.21 of the ATNF Pulsar Catalogue.  For comparison, the energy
flux of HESS\,J1303-631 between 0.3 and 10\,TeV implied by its best-fit
spectrum is $2.1 \times 10^{-11}$\,erg\,cm$^{-2}$\,s$^{-1}$.  Some
caveats should be kept in mind when comparing this number with the
listed values of $F_{\mathrm{sd}}$: uncertainties in the estimation of
$D$ for individual pulsars can be significant; the comparison also
implicitly assumes that the $\gamma$-ray emission is isotropic, but
this is almost certainly a valid assumption for an extended source
such as HESS\,J1303-631.  Finally, delayed energy release of particles
injected in the past, when the spindown luminosity was higher, could in
principle increase the apparent efficiency; but this is unlikely to be
a large effect in most evolutionary scenarios.

With this in mind, it can be seen that most of the pulsars listed
in Table~\ref{PulsarParameters} can be ruled out as possible counterparts
for HESS\,J1303-631 due to insufficient energetics.  Only PSR\,J1301-6305
has enough energy to power the whole of the VHE emission, and even in
this case the conversion efficiency would have to be unusually high.

\begin{figure}
 \centering
  \resizebox{\hsize}{!}{\includegraphics{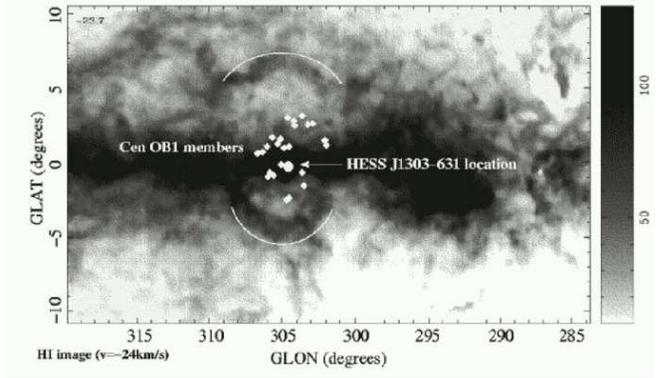}}
 \caption{\ion{H}{i} image ($\sim 35\degr \times 20\degr$) at $v=-24$ km 
s$^{-1}$ of the shell GSH~305+01$-$24 (McClure-Griffiths et al. 
\cite{Mclure:1}). The member stars of Cen~OB1, location of HESS\,J1303-631 
and approximate outer boundaries of GSH~305+01$-$24 are indicated.}
 \label{fig:HIshell}
\end{figure}

\subsection{Molecular clouds, energetic stellar winds from Cen~OB1, and SNR association}

If HESS\,J1303-631 is a result of cosmic ray interactions with ambient
gas, one can draw the following conclusions: the fact that the spectrum of
this source is significantly harder compared to the $E^{-2.7}$ diffuse
(local) galactic cosmic ray spectrum, indicates that one does not see an
interaction of this galactic cosmic ray component with regions of enhanced
ambient target gas density, but rather of a source of cosmic rays, which
is located within or near HESS\,J1303-631. If such a source is also in
contact with a dense molecular cloud, the increased target density results
in an increased rate for the production of VHE $\gamma$-rays as discussed
by Aharonian et al. (\cite{Aharonian1990}).

Molecular clouds can provide the conditions for both Type I and Type II
supernova events to occur, with the former resulting from accelerated
accretion induced collapse of white dwarfs due to the increased gas
density, and the latter resulting from the evolution of massive stars in
star formation regions towards supernovae.  Another possibility is that
stellar winds from OB stellar associations, interacting with molecular
clouds, may also be sources of $\gamma$-ray production, as suggested for
TeV\,J2032+4130 by Aharonian et al. (\cite{HEGRA_GalPlaneScan}), or, such
stellar winds may provide the required injection for further acceleration
in a nearby SNR shell from the same or another nearby OB association
(Montmerle \cite{Montmerle}), called supernova-OB associations (or SNOBs).
There is a possible correlation between SNOBs and unidentified
$\gamma$-ray sources as found by Montmerle (\cite{Montmerle}) in the COS-B
data, and in the EGRET data by Kaaret, \& Cottam (\cite{KC}).

As a first step the nature of molecular clouds in the direction of
HESS\,J1303-631 is investigated, as reflected by CO transition line
measurements, which gives (i) a kinematic distance based on galactic
rotation curves (sometimes one will get more than one solution), and, 
(ii) the column density of gas corresponding to the region from where 
the CO emissions are seen.

It is possible to fit the observed spectral shape of HESS\,J1303-631 with a hadronic model. A hadron
spectral index of $2.0$ combined with an exponential cut-off at $20$ to $50\,\mathrm{TeV}$ would even
account for the slight curvature apparent in the measured spectrum. To reproduce the observed TeV
flux, one has to assume a total cosmic ray energy of $E_{\mathrm{CR}} = 2$ to $5 \times 10^{49} \,
d_{\mathrm{kpc}}^{2} /\, n_{1} \, \mathrm{erg}$, where $d_{\mathrm{kpc}} = d / 1 \, \mathrm{kpc}$ is
the distance and $n_{1} = n / 1 \, \mathrm{cm}^{-3}$ is the associated target density.

The procedure of deriving $n$ from the velocity integrated CO temperatures
$\langle W(\mathrm{CO}) \rangle = \int T \mathrm{d}v$ is well-known,
provided that a scale size for the cloud is specified (see e.g. Dame et
al. (1993) and references therein). The deep CO survey of molecular clouds
in the southern Milky Way by Bronfman et al. (1989) covers the $300\degr$
to $348\degr$ longitude and $-2\degr$ to $+2\degr$ galactic latitude
ranges, with a spatial resolution of $0.125\degr$, which is smaller than
the resolved size of HESS\,J1303-631.

Extracting CO temperatures $T$ centred at $l=304.24\degr$ and
$b=-0.36\degr$ (the centroid for HESS\,J1303-631) from this survey,
results in three distinct spatial coincidences in velocity space. The
merits of these will be discussed below.

(i) It is unlikely that the Coalsack ($v=-4$ to $-2 \, \mathrm{km} \,
\mathrm{s}^{-1}$), at a distance of $\sim 175$ pc (Nyman et al.
\cite{Nyman:1}), is a candidate environment for HESS\,J1303-631 based on
two aspects: there is no evidence of star formation in this dark cloud
(Nyman et al. \cite{Nyman:1}), and a bright counterpart should have been
detectable at longer wavelengths, given the proximity of this cloud.

(ii) For the second CO solution the reader is refered to Figure (11) of
Bronfman (\cite{Bronfman:1}), who shows a clump in the velocity interval
$\sim -30$ to $-15 \, \mathrm{km} \, \mathrm{s}^{-1}$ along the
line-of-sight cutting through the centroid of the TeV source. The
temperature was integrated along this velocity interval to give
$W(\mathrm{CO})$, showing a plateau at the location of HESS\,J1303-631.
This quantity, if averaged over the surface of the resolved TeV source,
gives an average value of $\langle W(\mathrm{CO}) \rangle \sim 15$ K
$\mathrm{km} \, \mathrm{s}^{-1}$ for an average velocity of $v_0\sim -23$
$\mathrm{km} \, \mathrm{s}^{-1}$, which corresponds to a kinematic
distance of either $d=2.1$ or 7.7 kpc, assuming the kinematic
velocity model of Wouterloot \& Brand (\cite{Wouterloot:1}), with the
ambiguity arising from the galactic rotation curve.  The $1\degr$ cloud
diameter converts to densities of $n\sim 56$ cm$^{-3}$ and $n\sim 15$
cm$^{-3}$ for the two distances, respectively, which translates to a total
required cosmic ray energy of either $\sim 3\times 10^{48}$ (for $d=2.1$
kpc), or, $\sim 10^{50}$ ergs for $d=7.7$ kpc. The size of the
$\gamma$-ray source is determined by the size of the dense target (density
profile) and the diffusion (propagation) length. The densities derived
here will be required for the discussion below.

(iii) The third CO solution involves the detection of the giant molecular
cloud (GMC) G303.9-0.4 in the Carina arm at a distance of 12 kpc. The
$l-b$ CO map of G303.9-0.4 (integrated over velocities in the 7 to 50
$\mathrm{km} \, \mathrm{s}^{-1}$ interval) can be found in Grabelsky et
al. (1988). Two adjacent CO maxima are visible, with the TeV source
located on one of them, giving a TeV source averaged CO intensity of
$\langle W(\mathrm{CO})  \rangle \sim 8$ K $\mathrm{km} \,
\mathrm{s}^{-1}$. The target density at the contours corresponding to the
TeV maximum is then $n\sim 10$ to 20 $\mathrm{cm}^{-3}$, if a cloud scale
size corresponding to the FWHM of the integrated temperatures on the image
is assumed. The required energy in cosmic rays is now $E_{\mathrm CR}\sim
3\times 10^{50}$ ergs. Such energetics would require a relative powerful
supernova explosion and it remains to be shown if an old SNR shell with 40
pc radius can still support cosmic ray production at the required level.

An interesting counterpart for HESS\,J1303-631 in connection with CO
solution (ii) may be the OB stellar association Cen~OB1. Via the stellar
winds of their member stars, OB associations are thought to supply
considerable kinetic energy into the surrounding media. Cen~OB1 contains
$\sim$20 stars of spectral type O9, suggesting that some member stars may
have evolved beyond the supernova stage. The Wolf-Rayet star $\theta$-Mus
is also a member and may be a triple (WR + O + O stars) system (Hartkopf
et al. \cite{Hartkopf:1}). McClure-Griffiths et al. (\cite{Mclure:1})
identify the \ion{H}{i} shell GSH~305+01$-$24 as being blown out by the OB
association Cen~OB1. The estimated distance of GSH~305+01$-$24 at 2.2 kpc
is consistent with that of Cen~OB1 ($2.5 \, \mathrm{kpc}$, Humphreys
\cite{Humphreys:1}).  Moreover, the estimated kinetic energy required to
maintain the \ion{H}{i} shell expansion at $\sim 10\times10^{51}$ ergs is
met by that of Cen~OB1 at $\sim 8 \times 10^{51} \, \mathrm{ergs}$, based
on stellar wind luminosities and ages of member stars.
Figure~\ref{fig:HIshell} depicts the positional relationship between the
\ion{H}{i} shell, Cen~OB1 member stars and HESS\,J1303-631. Assuming a
canonical factor 0.1 for kinetic energy to comsic-ray energy conversion,
$\sim 8 \times 10^{50} \, \mathrm{ergs}$ would be available for cosmic-ray
production. This link between a TeV $\gamma$-ray source and an OB
association mirrors very closely that of the first unidentified TeV source
TeV~J2032+4130 and Cygnus~OB2 discovered by the HEGRA collaboration
(Aharonian et al. \cite{Aharonian:Cyg1, Aharonian:Cyg2}).

Note, that the distances of Cen~OB1 and GSH~305+01$-$24 coincide with the
first kinematic CO solution of 2.1 kpc (within uncertainties). A number of
scenarios discussed below could give rise to the TeV emission:

(a) The stellar winds from member stars of Cen~OB1 provide conditions
for the acceleration of cosmic rays, perhaps via shocks setup at the
interfaces of the various stellar winds. These cosmic rays in turn
could illuminate a nearby molecular cloud. If the average distance of
HESS\,J1303-631 to the Cen~OB1 member stars is of order $\leq
2^\circ$, the required energy to account for the TeV source, assuming
all member stars contribute, would be $4\times10^{48} (2.0/0.32)^2\,
\mathrm{erg}$ = $2\times10^{50}\, \mathrm{erg}$.  This is consistent
with the cosmic-ray energetics of Cen~OB1 discussed earlier. Here, a
source radius of 0.32$^\circ$ (95\% or 2$\sigma$ containment radius)
is assumed. A dense cloud situated at the location of HESS\,J1303-631,
could therefore be a valid solution. The energetics could also be met
by a single or a few member stars, provided they were close enough.  
The CO data however, also suggest the presence of numerous dense
clouds in the vicinity of Cen~OB1. The fact that TeV emission from
just one location is seen could imply that HESS\,J1303-631 is somehow
special in relation to the distance to the local cosmic-ray
accelerator, and its geometry. Detailed modeling of the acceleration
and diffusion of cosmic-rays in this region will be needed to verify
this scenario.

(b) As for situation (a) but the TeV emission may arise from a leptonic
  process such as inverse-Compton upscattering of local soft photons.
  The localised nature of HESS\,J1303-631 would then require some extra
  conditions namely: the soft photon distribution (which may be dominated by UV photons
  from the member stars) is also localised; the acceleration and
  diffusion properties of electrons is such that the energetic electrons are
  confined. 

(c) If the oldest member stars already evolved into the supernova stage, the
SNOB scenario of Montmerle (\cite{Montmerle}) may also explain the
existence of HESS\,J1303-631, but it remains then to be shown why this 
SNR is not seen at X-ray and radio wavelengths.

\subsection{Clumps of annihilating dark matter}

A possible source of TeV $\gamma$-ray emission could also be a clump of
annihilating dark matter particles (Moore et al. \cite{DarkMatter}) which
would result in an energy spectrum with a cut-off at the mass of the dark
matter particle. The galactic halo could contain a huge amount of such
clumps comprised e.g. of neutralinos.

Although not very likely, it is possible to reproduce in Monte Carlo
simulations the observed flux with the annihilation radiation from
high mass neutralinos, but it is not possible to reproduce the luminal
profile measured from HESS J1303-631 assuming a Navarro-Frenk-White
(NFW) dark matter density profile (Navarro et al. \cite{NFW_Profile}).
For the NFW and other commonly used profiles a narrower luminal
profile is expected than obtained from HESS\,J1303-631.

\section{Summary \& Conclusions}

A new unidentified TeV $\gamma$-ray source HESS\,J1303-631 was serendipitously
discovered in a dataset which was initially taken on the binary system
PSR\,B1259-63/SS\,2883 which was also discovered at TeV energies (see a parallel
paper: Aharonian et al. \cite{HESS_PSRB1259_Detection}). For the first time in TeV
$\gamma$-ray astronomy, the detection and analysis of two sources within the same
field of view is achieved which subsequently also occured in other pointings
towards the galactic plane, showing the potential of the new generation of
ground-based experiments -- such as H.E.S.S. -- with the stereoscopic observation
mode and its large field of view of $\sim 5\degr$.

HESS\,J1303-631 was found to be clearly extended with a width of an
assumed intrinsic Gaussian emission profile of $\sigma = (0.16 \pm 0.02)
\, \degr$. The energy spectrum can be described by a power-law with a
photon index of $\Gamma = 2.44 \pm 0.05_{\mathrm{stat}} \pm
0.2_{\mathrm{syst}}$. The integral flux above $380 \, \mathrm{GeV}$ was
found to remain on a constant level of $(17 \pm 3) \, \%$ of the flux from
the Crab Nebula during the observations taken between February and June,
2004.  Detailed studies on morphology and spatially-resolved energy
spectra are underway.

Up to now, no counterpart at other wavelengths was identified. However,
the location close to the galactic plane places HESS\,J1303-631 in the
vicinity of a variety of possible objects which might be involved in the
production mechanisms explaining the observed TeV $\gamma$-ray emission. A
pulsar wind nebula powered by the nearby pulsar PSR\,J1301-6305 is one of
the interesting candidates. Another possibility would be the Cen~OB1
stellar association at a distance of $2.5 \, \mathrm{kpc}$ which provides
enough kinetic energy by stellar winds. The interaction of an expanding
supernova remnant shell with a molecular cloud as target material can also
be seen as an interesting configuration for TeV $\gamma$-ray production.
Three molecular clouds are located along the line of sight of
HESS\,J1303-631: The Coalsack nebula at a distance of $\sim 175$ pc, a
cloud in the Carina-Sagittarius arm at $2.1 \, \mathrm{kpc}$, resp. $7.7
\, \mathrm{kpc}$, and the giant molecular cloud GMC G303.0-0.4 in the
Carina Arm at a distance of $12 \, \mathrm{kpc}$. Finally, the possibility
of a clump of annihilating dark matter particles was considered, which
however can be excluded since the luminal profile of HESS\,J1303-631 is
not reproduced.

In summary, no clear counterpart was identified to date. Therefore HESS\,J1303-631
has to be considered as the second unidentified TeV source detected, following
\object{TEV\,J2032+4130} discovered in the Cygnus region by HEGRA (Aharonian et al.
\cite{Aharonian:Cyg1, Aharonian:Cyg2}). Meanwhile, the results of the H.E.S.S.
galactic plane scan revealed more unidentified TeV $\gamma$-ray sources (Aharonian
et al. \cite{HESS_GalPlaneScan}), thus further opening the door to a new class of
(yet-unidentified) TeV $\gamma$-ray sources. To further investigate possible
production mechanisms and to understand this new region of the non-thermal universe
future multi-wavelength observations (especially in X-rays) are essential and
partially already initiated (Mukherjee \& Halpern \cite{HESSJ1303-Chandra}).

\begin{acknowledgements}

The support of the Namibian authorities and of the University of Namibia
in facilitating the construction and operation of H.E.S.S. is gratefully
acknowledged, as is the support by the German Ministry for Education and
Research (BMBF), the Max Planck Society, the French Ministry for Research,
the CNRS-IN2P3 and the Astroparticle Interdisciplinary Programme of the
CNRS, the U.K. Particle Physics and Astronomy Research Council (PPARC),
the IPNP of the Charles University, the South African Department of
Science and Technology and National Research Foundation, and by the
University of Namibia. We appreciate the excellent work of the technical
support staff in Berlin, Durham, Hamburg, Heidelberg, Palaiseau, Paris,
Saclay, and in Namibia in the construction and operation of the equipment.
This research has made use of the SIMBAD database, operated at CDS,
Strasbourg, France.

\end{acknowledgements}

\end{document}